\documentclass[iop]{emulateapj}
\usepackage{natbib}
\usepackage{amsmath}
\usepackage{graphicx}
\usepackage{aas_macros}
\usepackage{latexsym}
\usepackage{longtable}
\usepackage{float}
\usepackage{tablefootnote}
\usepackage{threeparttable}

\shorttitle{Are fossil groups early-forming?}
\shortauthors{Kundert et al.}

\begin{document}
 
\title{Are fossil groups early-forming galaxy systems?}

\author{
A. Kundert\altaffilmark{1,$\star$},
E. D'Onghia\altaffilmark{1,$\dagger$},
and J. A. L. Aguerri\altaffilmark{2,3}
}
 
\altaffiltext{1}{Department of Astronomy, University of Wisconsin-Madison, 475 N. Charter St., Madison, WI 53706, USA}
\altaffiltext{2}{Instituto de Astrof\'{i}sica de Canarias, C/ V\'{i}a L\'{a}ctea s/n, E-38200 La Laguna, Tenerife, Spain}
\altaffiltext{3}{Departamento de Astrof\'{i}sica, Universidad de La Laguna, E-38205 La Laguna, Tenerife, Spain}
\email{$^{\star}$ E-mail: kundert@astro.wisc.edu}
\altaffiltext{$\dagger$}{Alfred P. Sloan Fellow}

\begin{abstract}
 
Using the Illustris cosmological simulation, we investigate the origin of fossil groups in the $M_{200}=10^{13}-10^{13.5}M_{\odot}/h$ mass regime. We examine the formation of the two primary features of fossil groups: the large magnitude gap between their two brightest galaxies, and their exceptionally luminous brightest group galaxy (BGG). For fossils and non-fossils identified at $z=0$, we find no difference in their halo mass assembly at early times, departing from previous studies. However, we do find a significant difference in the recent accretion history of fossil and non-fossil halos; in particular, fossil groups show a lack of recent accretion and have in majority assembled 80\% of their $M_{200}(z=0)$ mass before $z\sim 0.4$. For fossils, massive satellite galaxies accreted during this period have enough time to merge with the BGG by the present day, producing a more massive central galaxy; and, the lack of recent group accretion prevents replenishment of the bright satellite population, allowing for a large magnitude gap to develop within the past few Gyr. We thus find that the origin of the magnitude gap and overmassive BGG of fossils in Illustris depends on the recent accretion history of the groups and merger history of the BGG after their collapse at $z\sim1$. This indicates that selecting galaxy groups by their magnitude gap does not guarantee obtaining neither early-forming galaxy systems nor undisturbed central galaxies.

\end{abstract}
 
\keywords{galaxies: groups: general - galaxies: formation - galaxies: evolution}

 
\section{Introduction}
 
Fossil galaxy systems have long been thought to be dynamically evolved due to both a central galaxy that dominates the total optical luminosity of the group as well as a large difference in brightness between their two brightest satellites. \citet{barnes1989} proposed that within a Hubble time, satellites within a compact group will merge with the central galaxy due to dynamical friction, to produce a singular bright massive central galaxy in the center of a group-sized dark matter halo. And indeed, the first identification of one of such systems was made by \citet{ponman1994}, finding that the group RX-J1340.6+4018 was an apparently isolated early-type galaxy surrounded by a X-ray halo with similar luminosity as a galaxy group.
 
The first observational definition for fossil groups (FGs), proposed by \citet{jones2003}, selected galaxy systems with an X-ray luminosity exceeding $L_{\mathrm{X}}\geq 10^{42} h_{50}^{-2}$ erg s$^{-1}$, and a magnitude gap greater than 2 mags in the \textit{R}-band within half the projected virial radius. The $L_{\mathrm{X}}$ requirement was motivated to select group and cluster mass systems, and the 2 mag or greater magnitude gap criterion selected the most extreme end of the observed magnitude gap distribution. Furthermore, calculating the gap within half the virial radius ensured $L$* galaxies initially at this distance have had time to merge with the central galaxy within a Hubble time. Using this definition, fossil systems have been observed at all masses \citep[see][]{cypriano2006,mendesdeoliveira2006,aguerri2011,zarattini2014}. Additionally, the galaxy luminosity functions of fossils identified in this way indicate their galaxy population depends on their magnitude gap. In particular, FGs have been found to show luminosity functions with a fainter characteristic magnitude as well as a slightly shallower faint-end slope \citep{zarattini2015}, possibly due to a deficit of dwarf galaxies \citep{donghia2004}.
 
Initial observations by \citet{jones2000,jones2003} indicated fossil BGGs had experienced no major mergers within the past 4 Gyr, and, in combination with the idea that $L$* galaxies had merged with the central galaxy, suggested fossil groups had built up their mass at an early epoch. Thus it was expected that both the halo and the BGG of fossil groups were old, and these systems have been evolving passively since their formation to the present day. However, this picture of dynamically evolved fossil groups has become less clear as larger samples of fossil groups have been studied.
 
The BGGs of fossil systems are among the brightest and most massive galaxies in the Universe
\citep{mendezabreu2012}. In addition, there is a relation between the brightness of the central galaxies and the magnitude gap of the systems. Systems with larger magnitude gaps show brighter central galaxies \citep[e.g.,][]{zarattini2014}; and moreover, the fraction of optical luminosity contained in the central galaxies of fossil systems is larger than in non-fossils \citep{harrison2012,zarattini2014}. However, the BGGs in fossil systems follow the same scaling relations as non-fossil BGGs \citep{mendezabreu2012}; and, no differences between fossil and non-fossil BGGs have been found in works related with stellar population properties \citep[see e.g.][]{labarbera2009,trevisan2017}. Studies focused on spatially resolved stellar population parameters, such as age and metallicity gradients, confirm that the BGG population of fossil systems are not a homogeneous class of objects. Additionally, the mass of fossil BGGs has been growing through merger events active until recent cosmic epochs \citep[see e.g.][]{eigenthaler2013,proctor2014}. 
 
The scaling relations describing the intracluster medium (ICM) of fossil and non-fossil systems have been a matter of debate in the literature. Some works have claimed that fossil systems are different to non-fossils in their optical luminosity \citep{proctor2011,khosroshahi2014}, X-ray temperature \citep{khosroshahi2007}, or the central concentration parameter of the dark matter halo \citep[e.g.,][]{sun2004}. Nevertheless, these differences are not confirmed by studies of large samples of FGs \citep[see e.g.][]{voevodkin2010,harrison2012,girardi2014,kundert2015,pratt2016}.
 
The galaxy substructure of fossil groups has also been analyzed in several studies in the literature. The absence of galaxy substructure is considered an indication of the dynamical relaxation of the system, which would be expected if fossil groups are truly dynamically old. \citet{aguerri2011} analyzed one FG finding no significant galaxy substructure. Nevertheless, the study from \citet{zarattini2016} on a larger sample of fossils, found that FGs show similar amounts of galaxy substructure as non-FGs. In addition, no differences have been found on the large-scale structure around fossil systems. Thus, some of them appear to be isolated structures, while in contrast others are embedded in denser environments \citep[see e.g.][]{adami2007,adami2012,pierini2011}.
 
In the present work we analyze the properties of fossils identified in the Illustris cosmological simulation. Our aim is to examine the properties of the groups selected from the simulation as a function of their magnitude gap in order to understand their dynamical evolution. We will focus our study on the the evolution of the magnitude gap of the systems, the formation and evolution of the BGGs, and the mass assembly history of the group halos.
 
The structure of this paper is as follows: Section~\ref{sec:simulation} is a brief overview of the Illustris simulation, with our sample of selected groups described in Section~\ref{sec:sample}. The results are presented in Section~\ref{sec:results} including an examination of the evolution of the magnitude gap (Sec.~\ref{sec:m12evolution}), and the properties of the brightest group galaxies (Sec.~\ref{sec:bgg}) and the group halos (Sec.~\ref{sec:groupma}). The discussion and conclusions of the paper are given in Sections~\ref{sec:discussion} and~\ref{sec:conclusions}, respectively.


\section{The Illustris Simulation}\label{sec:simulation}
 
The Illustris Project comprises a suite of cosmological simulations of varying resolution with hydrodynamics performed on a moving-mesh using AREPO \citep{springel2010a}. For our analysis of the evolution of fossil groups we make use of Illustris-1, the highest resolution simulation containing baryons in the Illustris suite. Illustris-1 contains $1820^3$ dark matter (DM) particles of mass $4.4*10^{6} M_{\odot}/h$, and initially $1820^3$ gas cells of average mass $8.9*10^{5} M_{\odot}/h$. The gravitational softening length for DM particles is 1.42 co-moving kpc for the duration of the simulation. For stellar particles, the gravitational softening length is 0.71 kpc at $z=0$, and fixed to the DM softening length at $z\geq1$. Gas cells and DM, stellar, and black hole particles are evolved within a periodic box of side length 75 co-moving Mpc/$h$ from initial cosmological conditions at $z=127$ to $z=0$, with 136 snapshots recorded between $z=47$ and the present.

The full-physics galaxy formation model of Illustris includes subgrid prescriptions for star formation and evolution; gas chemical enrichment with cooling and heating; black hole seeding and growth; and feedback from supernovae and AGN. Free-parameters in the feedback model were tuned to match observations, such as the evolution of the cosmic star formation rate density, in preliminary smaller volume test simulations \citep{vogelsberger2013}. The output galaxy population produced in Illustris reproduces a number of observations including the galaxy luminosity function at $z=0$ \citep{vogelsberger2014}, as well as the galaxy stellar mass function between $z=0-7$ \citep{genel2014}, in addition to others.
 
We utilize the halo and subhalo catalogues provided by the Illustris team \citep{nelson2015}, with relevant properties described here in brief. Group halos have been identified in the dark matter distribution using a friends-of-friends (FOF) algorithm \citep{davis1985} with a linking length 0.2 times the average interparticle separation. Within FOF group halos of at minimum 32 DM particles, gravitationally bound subhalos are identified from the total particle distribution using the SUBFIND algorithm \citep{springel2001,dolag2009}. Group centers and subhalo centers are set to be the coordinates of their most bound particle. The most massive subhalo in a group halo is considered the central subhalo. A FOF group's $R_{200}$ is defined to be the radius that encloses an average total particle density equal to 200 times the critical density. The $M_{200}$ of a group is calculated from the total mass of all baryons and dark matter enclosed within $R_{200}$.
 
SUBLINK merger trees have been constructed for Illustris by \citet{rodriguezgomez2015}. We trace the evolution of groups by following the main progenitor branch (MPB) of their $z=0$ central subhalos, and the properties of the groups these MPB subhalos inhabit at a given snapshot. For our analysis, we select groups at $z=0$ with a central subhalo whose MPB has been identified as centrals of their FOF groups at previous snapshots out to at least $z=0.1$. This ensures we are able to robustly track the evolution of the groups during the recent epoch.
 
We focus our analysis on subhalos with a total bound stellar mass exceeding log($M_{*}[M_{\odot}/h])\geq8$, for the purpose of requiring $\sim$100 stellar particles per galaxy. This cut in the stellar mass of subhalos produces a completeness limit in magnitude of $M_r=-16$ mag, as calculated from the summed luminosities of all bound subhalo stellar particles.
 
Illustris was run using WMAP-9 cosmological parameters \citep{hinshaw2013}. Complete details of the Illustris simulations and the galaxy formation model are described in \citet{vogelsberger2014} and \citet{vogelsberger2013}, respectively. Throughout this paper, we have made use of the publicly available online Illustris database \citep{nelson2015}.

 
\section{The Sample}\label{sec:sample}
 
\begin{table*}
\centering
\caption{Sample Properties}
\begin{threeparttable}
\begin{tabular}{l c c c c c c}
\hline
Subsample & $\Delta m_{12}$& $N_{\mathrm{groups}}$ & $ \log(M_{200})$& $ R_{200}$ [Mpc] & $N_{\mathrm{gal}}(R_{200})$ & $N_{\mathrm{gal}}(0.5R_{200})$ \\ [0.5ex]
\hline
nFG(0.5$R_{200}$) & 0-2 & 8 & 13.2$\pm$0.2 & 0.41$\pm$0.05 & 20$\pm$11.2 & 7$\pm$4.4 \\
FG(0.5$R_{200}$) & $\geq2$ & 46 & 13.3$\pm$0.2 & 0.42$\pm$0.05 & 19$\pm$9.1 & 7$\pm$3.4 \\
\hline
nFG($R_{200}$) & 0-2 & 25 & 13.2$\pm$0.2 & 0.42$\pm$0.05 & 20$\pm$8.3 & 7$\pm$3.9 \\
FG($R_{200}$) & $\geq2$ & 29 & 13.3$\pm$0.2 & 0.42$\pm$0.05 & 19$\pm$10.3 & 7$\pm$3.3 \\
\hline
\end{tabular}
\tablecomments{Properties of subsamples based on the magnitude gap within 0.5$R_{200}$ and $R_{200}$ for groups in the mass regime log($M_{200}[M_{\odot}/h]$)=13-13.5. $M_{200}$, $R_{200}$, and the number of galaxies, $N_{\mathrm{gal}}$, are average values for the subsample.}
\label{table:sample}
\end{threeparttable}
\end{table*}

\begin{figure}
\centering
\includegraphics[width=\linewidth]{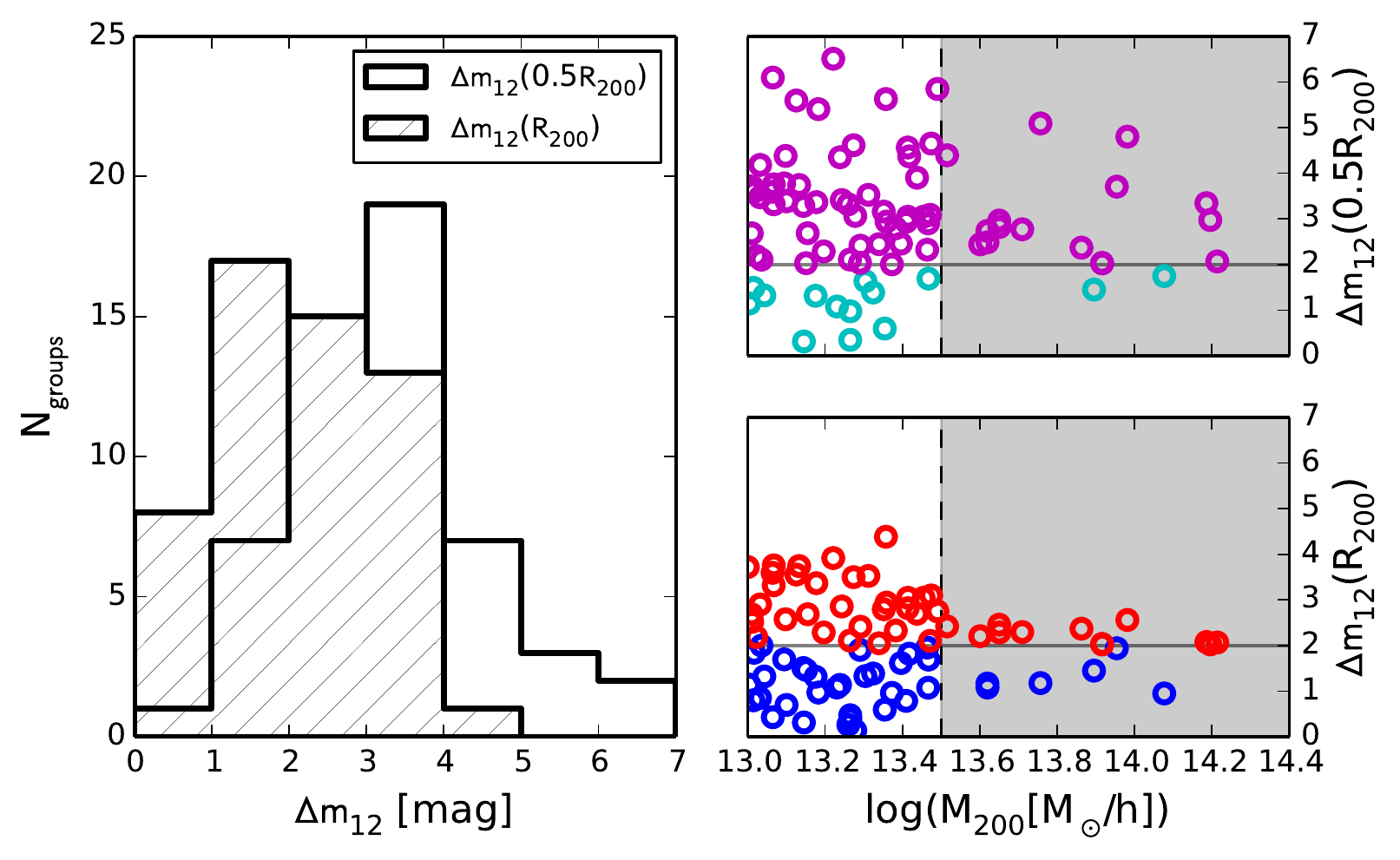} 
\caption{\textit{Left:} Magnitude gap distribution for Illustris groups with log($M_{200}[M_{\odot}/h]$)=13-13.5; shown for $\Delta m_{12}$ calculated within 0.5$R_{200}$ (solid histogram) and within $R_{200}$ (hatched histogram). \textit{Right:} Relation between magnitude gap and $M_{200}$. The shaded region is not used in our analysis, but is shown here to demonstrate the overall trend of magnitude gap and mass. Cyan and magenta circles represent non-fossils and fossils, respectively, identified by their gap within 0.5$R_{200}$. Blue and red circles represent non-fossils and fossils, respectively, identified by their gap within $R_{200}$.}
\label{fig:m12hist}
\end{figure}

To understand the dynamical evolution of fossil groups, we examine the formation of the magnitude gap, $\Delta m_{12}$, defined by the difference in $r$-band brightness between the first-ranked, $m_{1}$, and second-ranked, $m_{2}$, galaxies. The division between fossils (FGs) and non-fossils (nFGs) has traditionally been set at a gap of 2 mags, where fossils have a gap of $\Delta m_{12}(\mathrm{FG}) \geq 2$, and non-fossils have a gap of $\Delta m_{12}(\mathrm{nFG})=0-2$. We examine the evolution of the gap determined within both a volume with radius 0.5$R_{200}$ centered around the BGG, and within a volume with radius $R_{200}$.
 
We restrict our analysis of the formation of the gap to groups with mass $M_{200}=10^{13}-10^{13.5} M_{\odot}/h$. Observationally, groups with this mass are expected to meet the $L_{\mathrm{X}}$ requirement of \citet{jones2003}, see e.g. \citet{eckmiller2011}, although perhaps not all groups of this mass will be virialized. We do not rely on the X-ray luminosity, as is used in observational studies, because in Illustris the gas mass fraction within the inner regions of groups in our mass regime has been found to be a factor of 3-10 times lower than compared to observations, as noted in \citet{genel2014}. The upper limit of our $M_{200}$ selection is set to ensure a large enough sample size of both fossils and non-fossils. In the right panel of Fig.~\ref{fig:m12hist}, we show the distribution of magnitude gap and $M_{200}$ calculated for all FOF groups with mass $M_{200}\geq10^{13} M_{\odot}/h$. As can be seen above $10^{13.5} M_{\odot}/h$, which we do not include in our analysis, there are few non-fossils available for comparison to fossils, particularly for fossils and non-fossils defined by their gap within 0.5$R_{200}$. Thus we find the mass regime $M_{200}=10^{13}-10^{13.5} M_{\odot}/h$ is best for both examining the formation of the gap in a narrow group mass regime, as well as for comparing a significant sample size of fossils and non-fossils.
 
Our final sample of groups ($M_{200}=10^{13}-10^{13.5} M_{\odot}/h$) identified at $z=0$ consists of 46 FG(0.5$R_{200}$) and 8 nFG(0.5$R_{200}$), defined by their gap within 0.5$R_{200}$; as well as 29 FG($R_{200}$) and 25 nFG($R_{200}$), defined by their gap within $R_{200}$. The properties of these subsamples are recorded in Table~\ref{table:sample}, including average group $M_{200}$ mass, $R_{200}$ radius, and number of member galaxies.
 
Fig.~\ref{fig:m12hist}, left panel, shows the distribution of the magnitude gap values calculated for our sample of groups. For the gap calculated within $R_{200}$, we find a peak at $\Delta m_{12}(R_{200})\sim 1.5$, with 54\% of groups classified as FG($R_{200}$). Analytically, as predicted by the Press-Schechter formalism, 5-40\% of groups are expected to have magnitude gaps larger than 2 mags as calculated in a 500kpc projected radius, with the distribution peak at $\Delta m_{12}\sim1$ for groups of mass log(M)=13.5 \citep{milosavljevic2006}; we thus find a comparable magnitude gap distribution within $R_{200}$ as has been predicted. However, we find the peak of the gap distribution defined within 0.5$R_{200}$ occurs at $\Delta m_{12}(0.5R_{200})\sim$3, producing a relative abundance of 80\% fossils. In comparison, the FG($0.5R_{200}$) abundance is estimated from observations to be 8-20\% of groups with log($L_{\mathrm{X,bol}}[\mathrm{erg} \ \mathrm{s}^{-1}])\geq42$ \citep{jones2003}.
 
Differences between the Illustris gap distribution and those observationally found within 0.5$R_{200}$ might arise for a few reasons: (1) The central galaxies in Illustris are overmassive and overbright, compared to observations, as a result of the simulation feedback prescription \citep{vogelsberger2013,genel2014}; (2) Overmerging of satellite galaxies within the central regions of the groups is a known problem in simulations \citep{katzwhite1993}; (3) We do not employ the X-ray luminosity criterion that has been applied to previous observational studies. The overabundance of Illustris fossils, as compared to observations, has also been noted and discussed in \citet{raouf2016}.
 
We also see in Fig.~\ref{fig:m12hist}, a number of extreme gap, $\Delta m_{12}(0.5R_{200})\geq4$, groups exist in our FG(0.5$R_{200}$) subsample. These very large gap groups have a $m_{2}(z=0$) galaxy that is too faint to meet the completeness limit of many observational studies, and as a result these type of extreme gap objects have rarely been observed \citep[although see][]{zarattini2014}. However, 15/16 of these extreme gap groups have at higher redshift had a bright satellite pass within 0.5$R_{200}$, that has not merged, but has moved outside of this region at $z=0$ due to the path of its orbit. And indeed 15/46 of all FG(0.5$R_{200}$) have had a current satellite within 2 magnitudes of the brightness of the central galaxy pass within 0.5$R_{200}$ in the past without merging, indicating they would have been classified as non-fossil at other snapshots and are by chance currently identified as fossil in the present snapshot.
 
We also here note that the magnitude gap of an individual group is a highly transitory feature. The infall of new bright satellites accreted by the group will act to decrease the magnitude gap, while mergers of bright satellite galaxies with the BGG will cause the gap to increase. Additionally, the orbits of satellite galaxies within the group will produce a variance in the gap unrelated to mergers or infall, causing the gap to vary on short timescales particularly when the gap is calculated within half of $R_{200}$. The average maximum variance in $\Delta m_{12}$(0.5$R_{200}$) is $\sim$ 2 mag since $z\sim0.1$. While for $\Delta m_{12}$($R_{200}$) it is $\sim$ 0.7 mag. It is clear that with a variance of 2 mags within the past Gyr for groups defined by their $\Delta m_{12}$(0.5$R_{200}$) gap, there will also be a large variance in the properties of subsamples sorted by this metric at $z=0$.
 
Thus while the magnitude gap is traditionally defined within 0.5$R_{200}$ \citep{jones2003}, we find an overabundance of FG(0.5$R_{200}$) compared to observations and additionally find $\Delta m_{12}$(0.5$R_{200}$) is highly affected by the orbits of its satellites which obscures information about the dynamical state of groups characterized in this way. Most of our tests in our later analysis (Section~\ref{sec:results}) indeed show no statistical difference between FG(0.5$R_{200}$) and nFG(0.5$R_{200}$). On the other hand, while $\Delta m_{12}$($R_{200}$) is less typically used, the abundance of FG($R_{200}$) are in order with predictions and observations, and we would expect little effect due to orbits, and as a result a less transient gap characteristic. We therefore mostly rely on comparing groups divided into samples by their $R_{200}$ gap, to understand the physical processes driving the evolution of the magnitude gap. We will primarily depend on the results of this subsample for our understanding of the dynamical state of fossils, but also present the results within 0.5$R_{200}$ following observational convention.

\section{Results}\label{sec:results}
 
 
\subsection{Evolution of the magnitude gap}\label{sec:m12evolution}

\begin{figure}
\centering
\includegraphics[width=\linewidth]{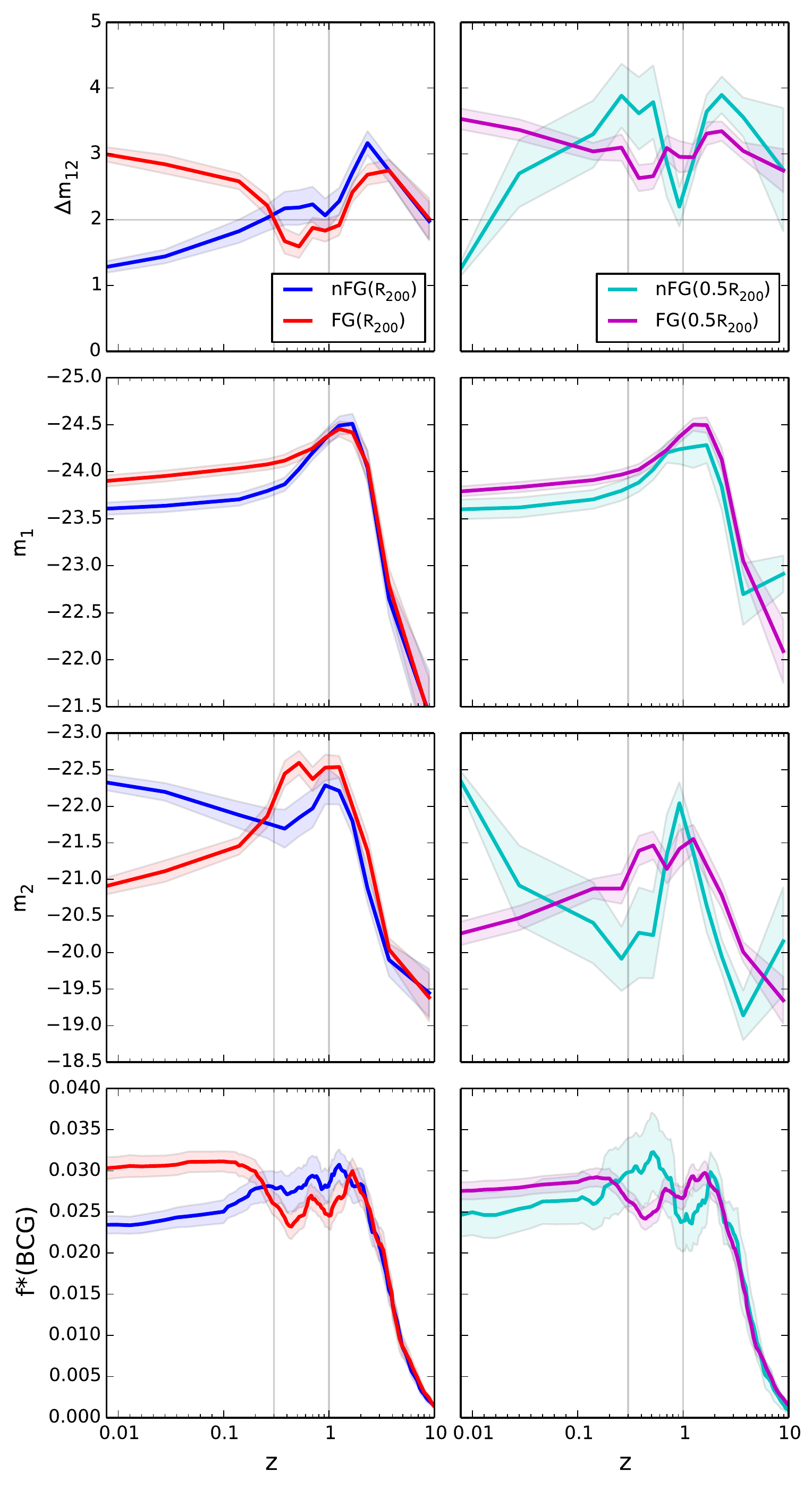} 
\caption{Average evolution of properties of fossils and non-fossils identified at $z=0$ and defined by $\Delta m_{12}$($R_{200}$) (\textit{left}) and $\Delta m_{12}$(0.5$R_{200}$) (\textit{right}).
\textit {First row:} magnitude gap calculated between first- and second-ranked galaxies at each redshift. Plotted values have been averaged over 1Gyr intervals to reduce scatter due to the high degree of transience of the gap.
\textit{Second row:} brightness of first-ranked galaxy identified at each redshift, $m_{1}(z)$.
\textit{Third row:} brightness of second-ranked galaxy identified at each redshift, $m_{2}(z)$.
\textit{Fourth row:} ratio of the BGG stellar mass to group $M_{200}$ at each redshift, i.e. \textit{f*}(BGG)=$M_{*,\mathrm{BGG}}(z)$/$M_{200,\mathrm{group}}(z)$.
1$\sigma$ errors calculated from 1000 bootstrap resamplings are shown.
}
\label{fig:m12z}
\end{figure}

Fossil groups are characterized by both their large magnitude gap as well as an overluminous central galaxy \citep{jones2003}. In Fig.~\ref{fig:m12z} we investigate how the evolution of these two characteristics are related. In the upper row of Fig.~\ref{fig:m12z}, we present the evolution of $\Delta m_{12}(z)$ within $R_{200}$ (left) and 0.5$R_{200}$ (right) for fossils and non-fossils defined by their gap at $z=0$. To further understand the $\Delta m_{12}(z)$ evolution, we also include in the middle rows the evolution of $m_{1}(z)$ and $m_{2}(z)$, the brightness of the first- and second-ranked galaxies identified at each redshift. Because the magnitude gap and $m_{2}(z)$ galaxy brightness are transitory properties for individual groups, particularly within $0.5R_{200}$, we display the magnitude gap and brightness of $m_{1}(z)$ and $m_{2}(z)$ averaged over a timespan of 1 Gyr in the first three rows. In the bottom row, we show the averaged fraction of a group's mass contained in the BGG's stellar component, which for convenience we refer to as \textit{f*}(BGG)=$M_{*,\mathrm{BGG}}$/$M_{200,\mathrm{group}}$. This \textit{f*}(BGG) quantity is useful for comparing the relative mass of the BGG for its halo mass. We divide these plots into three redshift regimes: $z\geq1$, $z=0.3-1$, and $z\leq0.3$, during which we observe different phases in the evolution of the groups.
 
Prior to $z\sim1$, all groups have similar magnitude gaps and \textit{f*}(BGG); this epoch marks the time period when these groups are still in the process of assembling the majority of their halo mass. After $z\leq1$, we can clearly see the evolution of the magnitude gap and the evolution of \textit{f*}(BGG) are related: groups with large magnitude gaps also have large \textit{f*}(BGG). Between $z\sim0.3-1$, groups identified as FG($z=0$) are shown to have a smaller gap and lower \textit{f*}(BGG) ratio than their nFG($z=0$) counterparts. After $z\sim0.3$, we see these trends reverse such that by the present day FG($z=0$) have a larger magnitude gap and greater \textit{f*}(BGG) than nFG($z=0$). We thus see that the characteristically large magnitude gap of fossils identified at $z=0$ has formed only in the past few Gyr, and furthermore both fossils and non-fossils have evolved in their magnitude gap and ratio of BGG mass to halo mass since $z\sim1$.
 
These trends support the idea of a `fossil phase' as proposed by \citet{vonbendabeckmann2008}; fossils identified at $z=0$ were non-fossils at high redshift, while $z=0$ non-fossils were previously fossils. Indeed we find all $z=0$ identified fossils have been previously non-fossils, while all $z=0$ non-fossils have been previously fossil. In general, the average FG($R_{200}$)($z=0$) was last non-fossil $\sim$3 Gyr ago, while nFG($R_{200}$)($z=0$) were on average fossil $\sim$2 Gyr ago. And importantly, we find the relative overmassiveness of the BGG followed with large magnitude gap systems, such that when $z=0$ non-fossils had a large magnitude gap between $z=0.3-1$, they also had a relatively more massive \textit{f*}(BGG). Thus we would expect a sample of fossils identified at higher redshifts would also have a more massive BGG than non-fossils identified at the same redshift, although the magnitude gap will likely evolve by the present day.

 
\subsection{Properties of the BGG}\label{sec:bgg}
 
To understand how fossils obtain their characteristic overmassive BGG, in this section we investigate BGG scaling relations, stellar mass assembly history, and merger history.

\subsubsection{Scaling relations}\label{sec:scalingrelations}

\begin{figure}
\centering
\includegraphics[width=1.\linewidth]{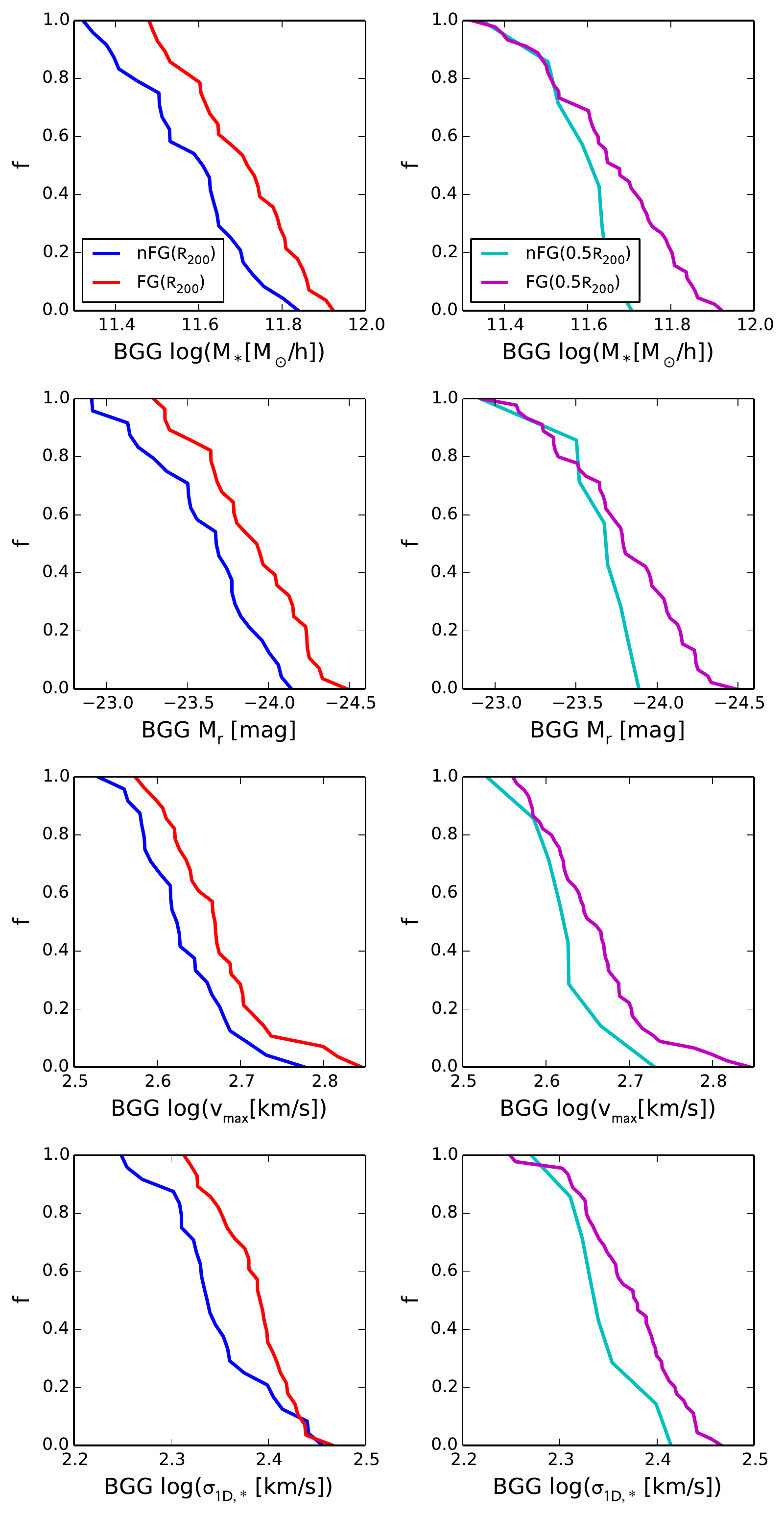} 
\caption{Distribution of BGG stellar mass, r-band magnitude, peak rotation curve circular velocity, and 1D velocity dispersion. Here the velocity dispersion has been calculated for the stellar particles within the stellar half-mass radius of the BGG.}
\label{fig:bcgprop}
\end{figure}
 
\begin{figure}
\centering
\includegraphics[width=1.\linewidth]{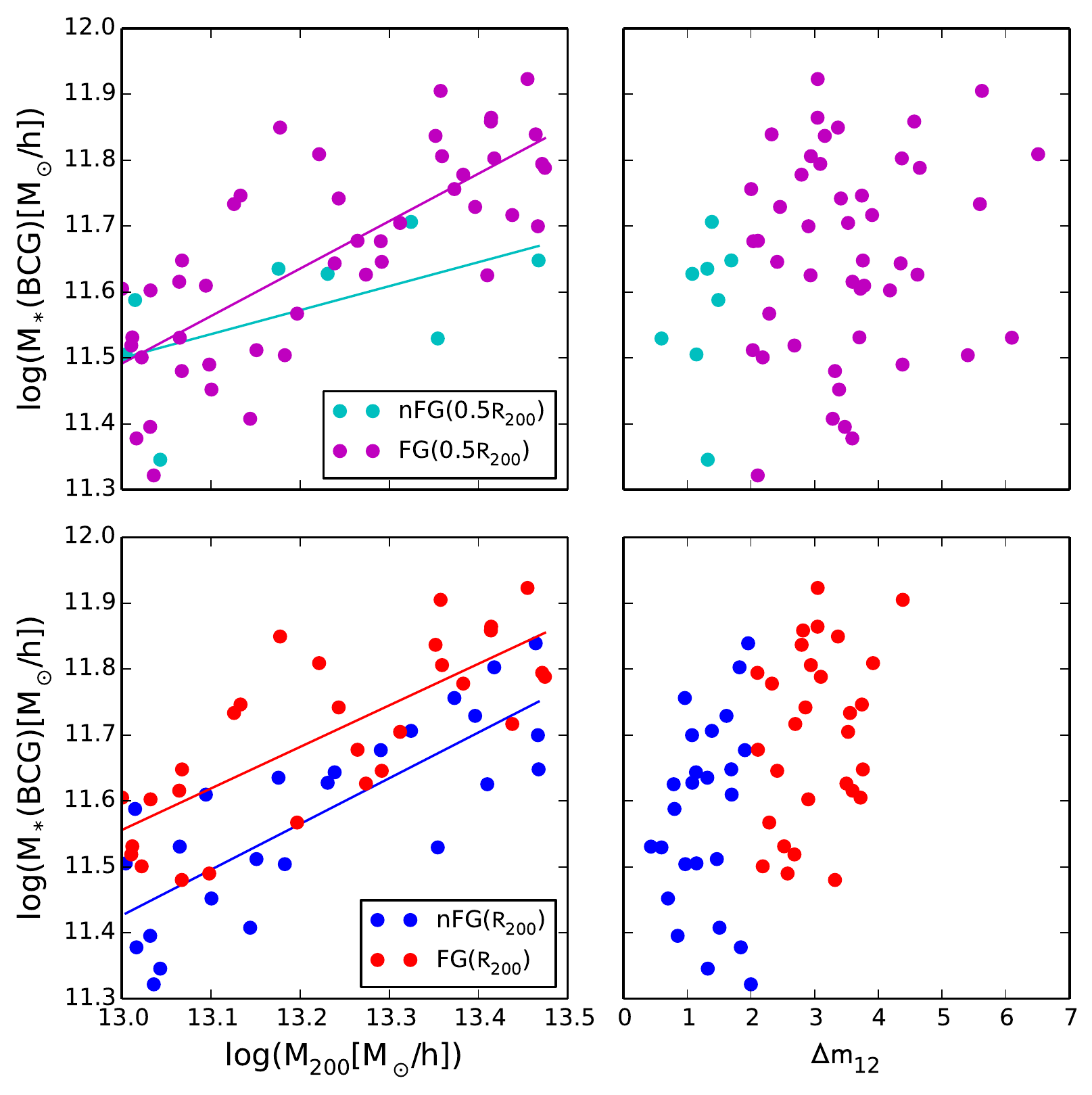} 
\caption{\textit{Left}: Stellar mass of BGG compared to the $M_{200}$ mass of the group in which it resides. Fossil($R_{200}$) BGGs are found to be more massive than non-fossil($R_{200}$) BGGs for the same group mass. \textit{Right}: A positive trend is found between the magnitude gap $\Delta m_{12}$($R_{200}$) and BGG stellar mass: the most massive central galaxies reside in the largest magnitude gap systems.}
\label{fig:sr_bcg}
\end{figure}

\begin{figure}
\centering
\includegraphics[width=1.\linewidth]{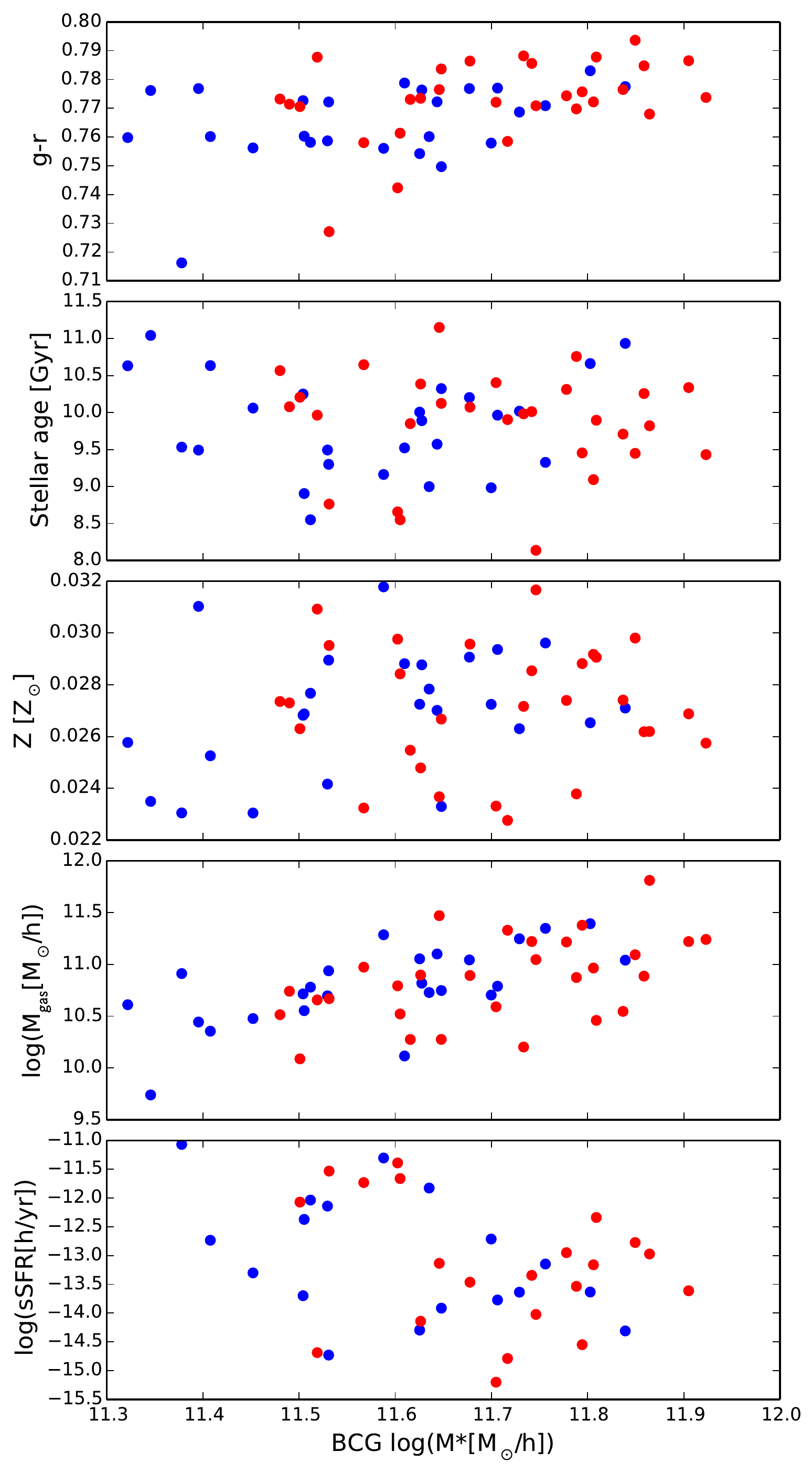}
\caption{Average mass-weighted $z=0$ properties for BGGs. Fossil($R_{200}$) BGGs, shown in red, and non-fossil($R_{200}$) BGGs, shown in blue, are similar in observable properties.}
\label{fig:m1propz0}
\end{figure}
 
Observationally, the central galaxies of fossil groups have been found to be more luminous and more massive than the central galaxies in non-fossil groups of the same global X-ray luminosity and temperature \citep{harrison2012,zarattini2014}. Furthermore, the centrals of fossil groups have been found to reside on the most massive end of the Faber-Jackson relation \citep{mendezabreu2012}. The group scaling relations of total optical luminosity ($L_{r}$) and bolometric X-ray luminosity ($L_{\mathrm{X}}$) are consistent for both fossil and normal groups \citep{harrison2012,girardi2014,kundert2015}, indicating a similar amount of baryonic mass. However a significant fraction of a fossil group's total optical luminosity is contributed by its central galaxy, suggesting, in combination with the $L_{r}$-$L_{\mathrm{X}}$ relations and large magnitude gap, that fossils have their stellar mass distributed differently than non-fossils.
 
Qualitatively matching these observational studies, we indeed find that the fossil BGGs in Illustris are more massive than non-fossil BGGs, and this characteristic is reflected in the distributions of BGG \textit{r}-band brightness, peak circular velocity, and central velocity dispersion (Fig.~\ref{fig:bcgprop}).
 
In Fig.~\ref{fig:sr_bcg}, left panel, we present the scaling relations of BGG stellar mass and group $M_{200}$, along with the least squares best-fit relation. It can be clearly seen that the BGGs of FG($R_{200}$) are more massive than the BGGs of nFG($R_{200}$) residing in halos of the same mass, qualitatively matching the observed scaling relations results. Furthermore, a two-sample Kolmogorov–-Smirnov (KS) test on the ratio of the BGG stellar mass to the group $M_{200}$, \textit{f*}(BGG)=$M_{*}$(BGG)/$M_{200}$(group), strongly indicates the \textit{f*}(BGG) of FG($R_{200}$) and nFG($R_{200}$) follow a different distribution ($p_{\mathrm{KS}}=0.001$). In Fig.~\ref{fig:sr_bcg}, right panel, we also find the magnitude gap and the BGG stellar mass are correlated for the gap calculated within $R_{200}$, (Spearman $\rho=0.45$, $p=0.0007$), with the largest gap groups possessing the most massive BGGs. This suggests the mechanisms which produce an overmassive FG($R_{200}$) BGG are related to the physical processes that produce the magnitude gap, in good agreement with observations \citep[e.g.,][]{harrison2012,zarattini2014}.
 
However, as also revealed in Fig.~\ref{fig:sr_bcg}, comparing fossils and non-fossils defined by their $\Delta m_{12}$(0.5$R_{200}$) gap does not produce statistically different results. A two-sample KS test of the ratio of BGG stellar mass and group $M_{200}$ for nFG(0.5$R_{200}$) and FG(0.5$R_{200}$) shows no difference in their distributions with $p_{\mathrm{KS}}=0.69$. And in testing for correlation between $\Delta m_{12}$(0.5$R_{200}$) and $M_{200}$, the Spearman test returns $\rho=0.18$, $p=0.19$. This further supports $\Delta m_{12}$(0.5$R_{200}$) is highly affected by random chance, i.e., the location of satellite galaxies along their orbital paths as discussed in Section~\ref{sec:m12evolution}. It will thus be difficult to disentangle the effects driving how FG(0.5$R_{200}$) BGGs become overmassive and overluminous.
 
Despite the large \textit{f*} of FG($R_{200}$) BGGs, in Fig.~\ref{fig:m1propz0} we find no obvious difference in the color, stellar age and metallicity, total gas mass, and specific star formation rate (sSFR) of the central galaxies of the same stellar mass. All BGGs have a log(sSFR[M$_{\odot}$/yr])$\leq$-11, which is typically considered to be quenched at $z=0$. As might be expected, no differences are found when comparing the observational properties of BGGs separated by $\Delta m_{12}$(0.5$R_{200}$).


\subsubsection{BGG stellar mass assembly history}\label{sec:bcgma}
 
\begin{figure}
\centering
\includegraphics[width=1.\linewidth]{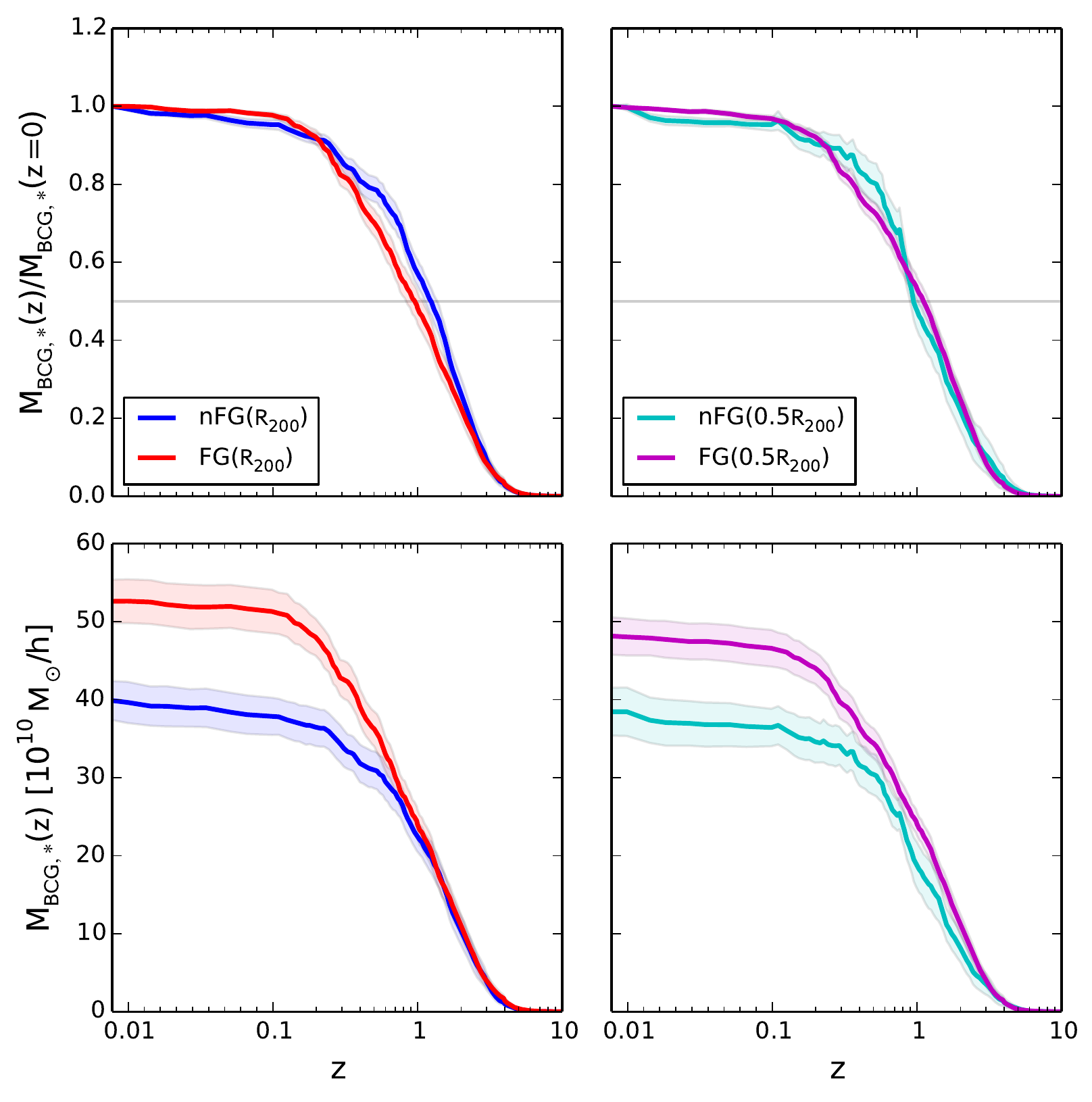} 
\caption{\textit{Top}: Average normalized stellar mass assembly history for central galaxies. \textit{Bottom}: Average stellar mass assembly history for central galaxies. 1$\sigma$ errors calculated from 1000 bootstrap resamplings are shown.}
\label{fig:bcgma}
\end{figure}

Given that fossil BGGs are overmassive and overluminous for their group $M_{200}$, we here examine how these galaxies build up their mass over time. In Fig.~\ref{fig:bcgma}, we show the average stellar mass assembly history (bottom) of the central galaxies, as well as the assembly history normalized by the final $z=0$ stellar mass (top).
 
The average evolution of the BGG stellar mass shows fossil central galaxies experience significant growth over the range $z\sim0.1-1$ relative to non-fossil BGGs. Indeed between $z=0-1$, FG($R_{200}$) BGGs increase in mass on average by a factor of $2.5\pm0.20$, compared to a factor of $1.9\pm0.14$ shown by nFG($R_{200}$) BGGs. However, we note that while fossil BGGs are on average more massive, they are still less than a factor of 2 more massive than the nFG BGGs at $z=0$. 
 
The normalized BGG stellar mass assembly history reveals FG($R_{200}$) BGGs have assembled 50\% of their final stellar mass by $z_{50}(\mathrm{BGG},*)=1.0\pm0.1$, while for nFG($R_{200}$) BGGs we find $z_{50}(\mathrm{BGG},*)=1.3\pm0.1$ (Table~\ref{table:bcgma}). Thus by this metric fossil BGGs have more recently assembled their stellar component, which is complementary to finding increased mass growth of the fossil BGGs after $z\sim1$. For FG(0.5$R_{200}$) and nFG(0.5$R_{200}$), we find no differences in the rate at which the stellar component is assembled, but do find FG(0.5$R_{200}$) BGGs grow more in mass after $z\sim1$ compared to nFG(0.5$R_{200}$), similar to our finding for FG($R_{200}$).

\begin{table}
\centering
\caption{BGG mass assembly}
\begin{threeparttable}
\begin{tabular}{l c c c c}
\hline
Subsample & $z_{50}$(BGG,*) & $\mathrm{N}_{\mathrm{major}}$ & $z_{\mathrm{LMM}}$ \\ [0.5ex]
\hline
nFG(0.5$R_{200}$) & 1.1 $\pm$ 0.1 & 2.9 $\pm$ 0.3 & 0.7 $\pm$ 0.1 \\
FG(0.5$R_{200}$) & 1.2 $\pm$ 0.1 & 3.2 $\pm$ 0.2 & 1.0 $\pm$ 0.2 \\
\hline
nFG($R_{200}$) & 1.3 $\pm$ 0.1 & 2.7 $\pm$ 0.2 & 1.1 $\pm$ 0.2 \\
FG($R_{200}$) & 1.0 $\pm$ 0.1 & 3.5 $\pm$ 0.3 & 0.8 $\pm$ 0.2 \\
\hline
\end{tabular}
\tablecomments{Average BGG stellar mass assembly time $z_{50}$(BGG,*), number of major mergers $\mathrm{N}_{\mathrm{major}}$, and redshift of last major merger $z_{\mathrm{LMM}}$. 1$\sigma$ errors have been bootstrapped.}
\label{table:bcgma}
\end{threeparttable}
\end{table}


\subsubsection{BGG merger history}\label{sec:bcgmerge}

\begin{figure}
\centering
\includegraphics[width=1\linewidth]{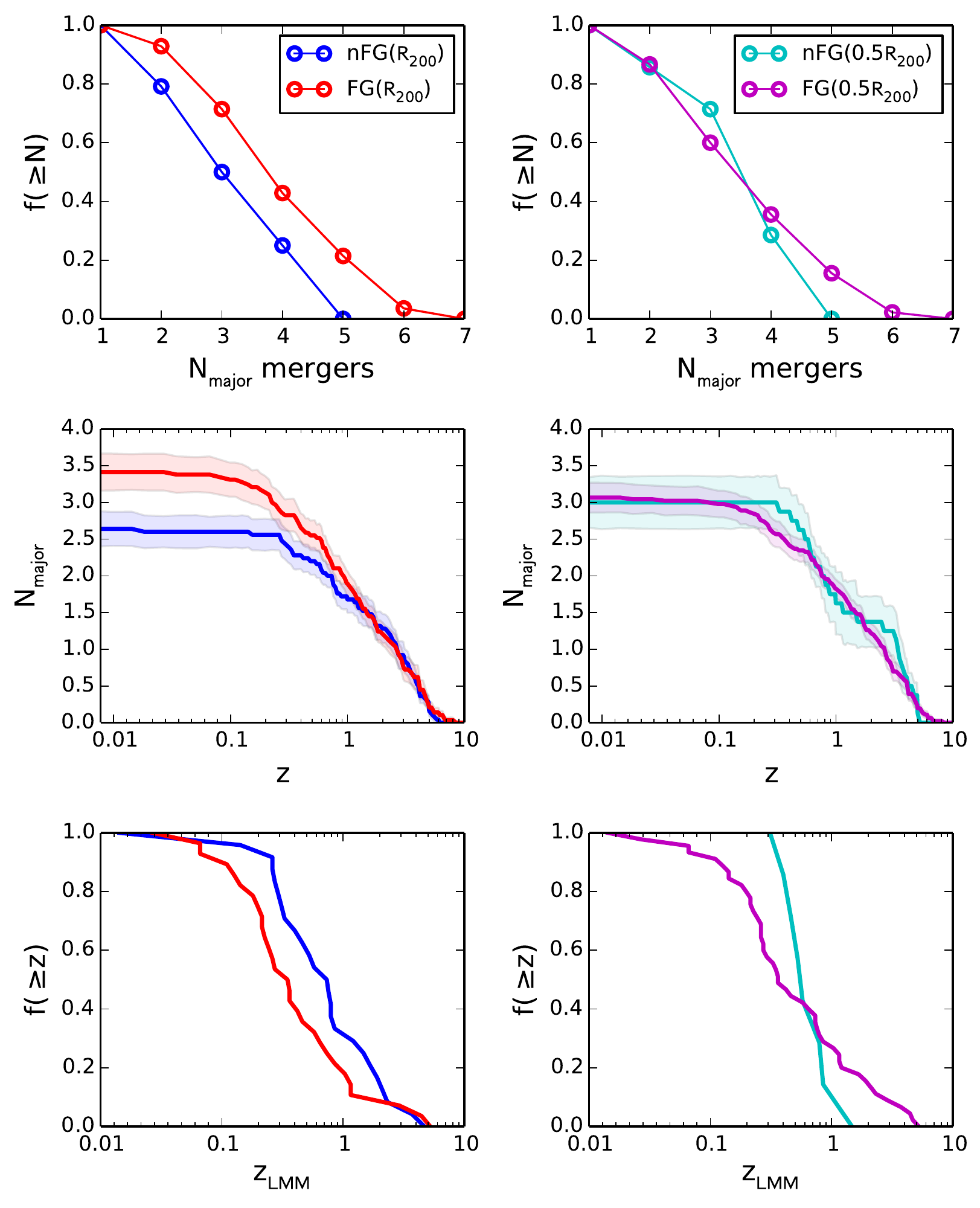} 
\caption{
\textit{Top row}: distribution of the total number of major mergers experienced by the BGGs at $z=0$.
\textit{Middle row}: average number of cumulative major mergers across $z$.
\textit{Bottom row}: distribution of the redshift of the last major merger of the BGGs.
}
\label{fig:nmergez}
\end{figure}

\begin{figure}
\centering
\includegraphics[width=1\linewidth]{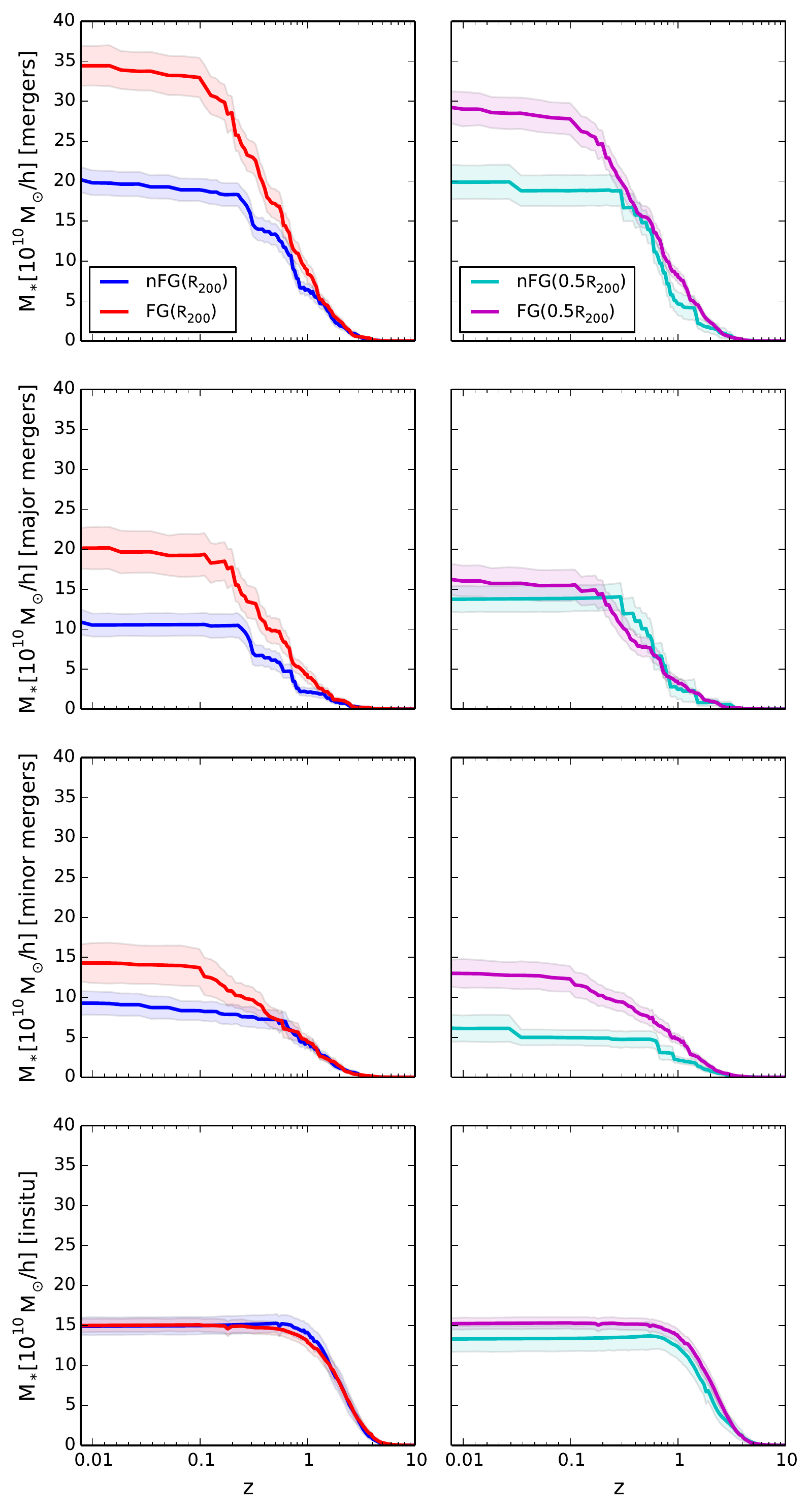} 
\caption{The average BGG stellar mass at each redshift originating from mergers (\textit{first row}), major mergers (\textit{second row}), minor mergers (\textit{third row}), and in situ formation (\textit{fourth row}). 1$\sigma$ errors from 1000 bootstrap resamplings are shown.}
\label{fig:smahz_mass}
\end{figure}

Differences in the stellar mass assembly history of fossil and non-fossil BGGs are likely to result from differences in their merger histories. In Fig.~\ref{fig:nmergez}, we show the major merger history of the BGGs, both in the evolution of the cumulative number of major mergers as well as the time of the last major merger. We here consider a major merger as a merger between galaxies with stellar mass ratios $M_{*,\mathrm{satellite}}/M_{*,\mathrm{BGG}} \geq 0.25$. Following \citet{rodriguezgomez2015}, the merger ratio between galaxies is calculated from the mass of both galaxies at the snapshot when the secondary galaxy is at its most massive. This avoids numerical effects and the transfer of mass shortly prior to when the merger of the galaxies occurs.
 
For FG($R_{200}$) and nFG($R_{200}$) BGGs, we find a difference in major merging history. Fig.~\ref{fig:nmergez} shows that FG($R_{200}$) BGGs will experience $\sim1$ more major merger than nFG($R_{200}$) BGGs. The distribution of the redshift of the last major merger, $z_{\mathrm{lmm}}$, is also shifted to more recent times for FG($R_{200}$) BGGs as also shown in Fig.~\ref{fig:nmergez}. Additionally, $\sim50$\% of FG($R_{200}$) BGGs experience 2 or more major mergers between $z=0-1$, compared to 20\% of nFG($R_{200}$) BGGs. No significant difference is found in the number of major mergers for FG(0.5$R_{200}$) and nFG(0.5$R_{200}$) BGGs.
 
In Fig.~\ref{fig:smahz_mass}, we examine the amount of mass the BGGs acquire through merging(major+minor), major merging, minor merging, and in situ star formation according to the Illustris stellar assembly catalog of \citet{rodriguezgomez2016}. In this catalog, star particles bound to $z=0$ galaxies are traced back and categorized by their origin location. For a given galaxy, in situ stars were formed from gas cells bound to the main progenitor branch of the galaxy, while star particles acquired from mergers were identified as having formed in a progenitor that has merged with the galaxy's main progenitor branch.

FG($R_{200}$) BGGs build up on average 64\% of their stellar mass through mergers (major+minor), compared to 50\% of mass acquired through mergers for nFG($R_{200}$) BGGs. This $\sim15$ percentage point difference is due to a $\sim 10$ percentage point greater contribution from major merging and a $\sim 5$ percentage point difference in minor merging for FG($R_{200}$) BGGs over nFG($R_{200}$) BGGs. On the other hand, in situ star formation contributes a similar amount of stellar mass for both FG($R_{200}$) and nFG($R_{200}$) BGGs. Thus mergers, and especially major mergers, seem primarily responsible for elevating the mass of FG($R_{200}$) BGGs over their nFG($R_{200}$) BGG counterparts between $z\sim0.1-1$.
 
The stellar mass contribution results for the groups defined by their $\Delta m_{12}(0.5R_{200}$) gap are less clear. Fig.~\ref{fig:smahz_mass} suggests the main difference between FG(0.5$R_{200}$) BGGs and nFG(0.5$R_{200}$) BGGs is in the mass acquired through minor merging, however we note that the FG(0.5$R_{200}$) BGGs and nFG(0.5$R_{200}$) BGGs also have similar masses for the same group $M_{200}$ (see Fig.~\ref{fig:sr_bcg}).
 
In summary, we have found $\Delta m_{12}$($R_{200}$) improves the identification of BGGs that are relatively overmassive for their group $M_{200}$. FG($R_{200}$) BGGs are statistically more massive and more luminous than nFG($R_{200}$) BGGs in group halos of the same $M_{200}$. FG($R_{200}$) BGGs assemble 50\% of their final stellar mass somewhat later than nFG($R_{200}$) BGGs, and additionally are more likely to have a more recent major merger. The larger \textit{f*} of FG($R_{200}$) BGGs is attributable to a greater amount of mass acquired through merging between $z=0.1-1$, with increased contribution from major merging providing the most significant boost to the mass of the fossil BGG compared to the non-fossil BGG. While we have shown that there are indeed statistical differences in how and when the stellar mass of FG($R_{200}$) and nFG($R_{200}$) BGGs is assembled, these differences do not produce any noticeable variations in observational properties such as color, stellar age, or sSFR.

 
\subsection{Group mass assembly history}\label{sec:groupma}
 
\begin{figure}
\begin{minipage}{\linewidth}
\centering
	\includegraphics[width=\linewidth]{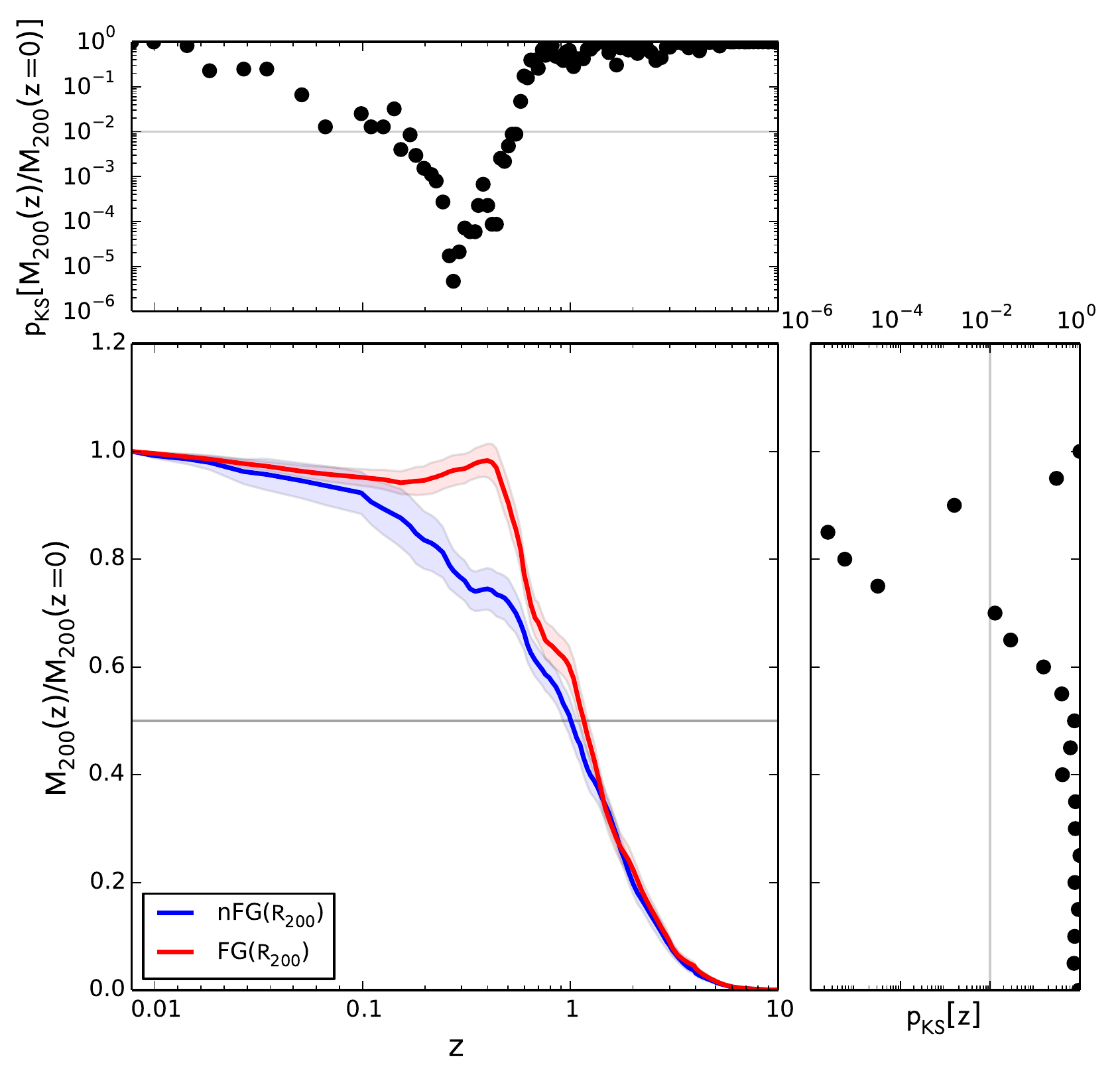} 
\end{minipage}
\begin{minipage}{\linewidth}
\centering
	\includegraphics[width=\linewidth]{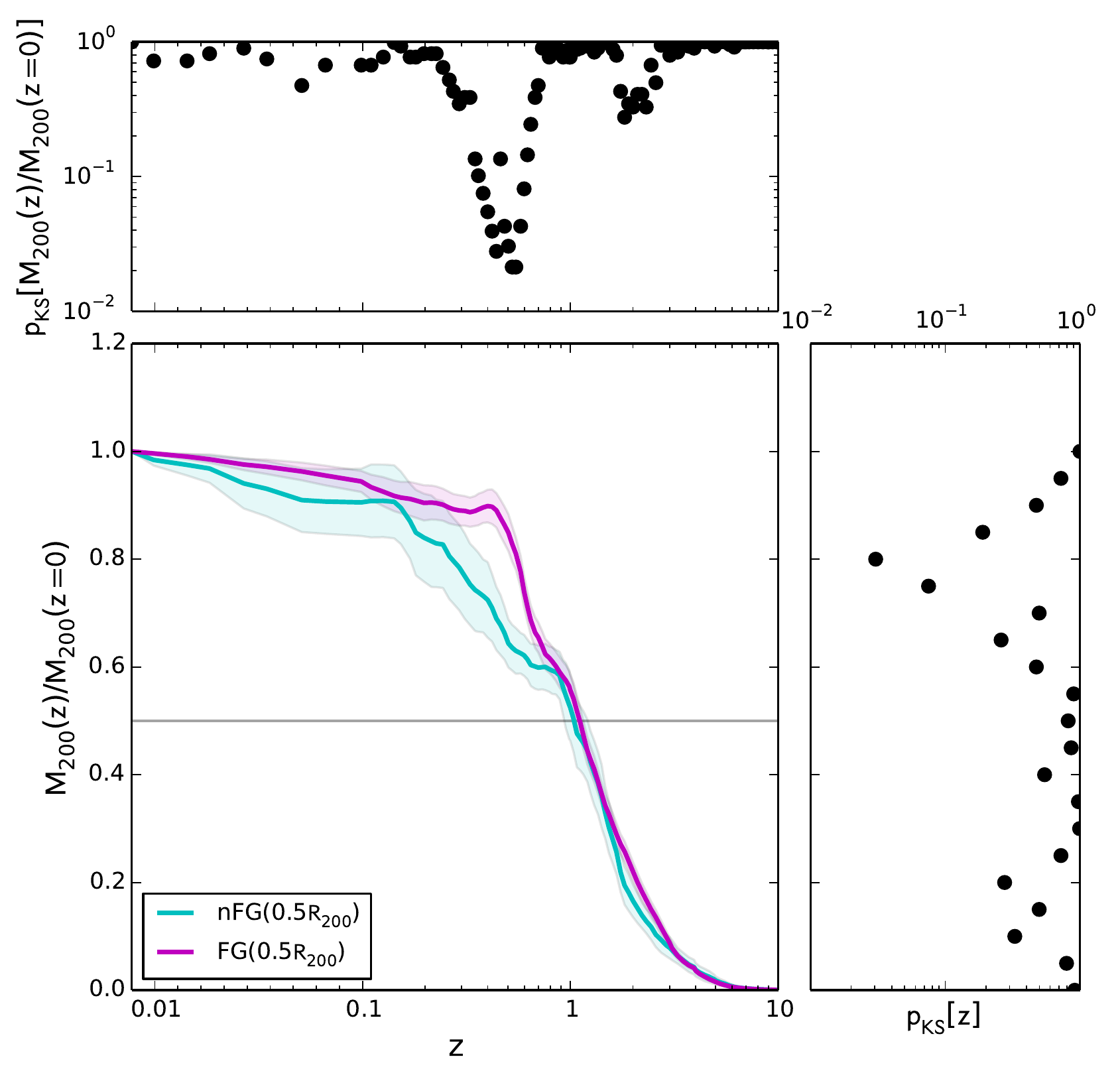} 
\end{minipage}
\caption{The average normalized group $M_{200}$ mass assembly history of fossils and non-fossils. 1$\sigma$ errors from 1000 bootstrap resamplings are shown.
\textit{Top x-axis side panel}: the two-sample KS test between non-fossils and fossils for the fraction of mass assembled at a given redshift.
\textit{Right y-axis side panel}: the two-sample KS test between non-fossils and fossils for the redshift at which some fraction of the final $M_{200}$ is assembled, with $p_{\mathrm{KS}}=0.01$ marked with a solid line. FG($R_{200}$) and nFG($R_{200}$) are different in their assembly epochs of $z_{70}$-$z_{90}$ with $p_{\mathrm{KS}}\leq0.01$, and are different ($p_{\mathrm{KS}}\leq0.01$) in the fraction of mass assembled between $\sim$2-5 Gyrs ago.}
\label{fig:groupma}
\end{figure}

Given the difference in the BGG mass assembly of fossils and non-fossils shown in the previous section, we might expect a difference in how the groups assemble their halo mass. Here we examine the mass assembly history of the groups. Fig.~\ref{fig:groupma} shows the average group $M_{200}$ assembly history normalized by the $M_{200}(z=0)$ mass. The mass assembly history of the groups was determined by tracing back the BGG($z=0$) main progenitor branch and using the associated group $M_{200}(z)$ at each redshift. Snapshots where the BGG($z=0$) main progenitor was not identified as the BGG($z=0$) of its group were excluded, and the missing $M_{200}$($z$) were estimated from linear interpolation. We compute a two-sample KS test on the fraction of mass assembled for the non-fossil and fossil sample at each redshift (top x-axis), as well as for the distribution in redshift at which a particular fraction of $M_{200}$($z=0$) is assembled (right y-axis). We find the mass assembly histories of fossils and non-fossils are similar at early times, but show an apparent divergence occurring after $z\sim1$.
 
In the literature, the redshift at which a halo builds up 50\% of its final $M_{200}(z=0)$ mass, $z_{50}$, has been frequently used as a metric for the formation time of the halo \citep[e.g.,][]{li2008}. And historically, previous fossil studies have particularly been interested in this time of halo assembly. We find that all groups in our sample on average form at $z_{50}\sim1$, although there is a wide range of $z_{50}$ times spanning $z_{50}=0.1-2$. Unlike previous studies of the mass assembly history of fossil groups \citep[e.g.,][]{dariush2007,diazgimenez2008,dariush2010}, we find no significant difference in the $z_{50}$ formation times with the KS test returning $p_{\mathrm{KS}} \sim 0.8$ for both FG($R_{200}$) and nFG($R_{200}$), as well as FG(0.5$R_{200}$) and nFG(0.5$R_{200}$).
 
However, the difference in the recent accretion history of fossils and non-fossils is particularly clear for $z_{80}$, the redshift at which these groups acquire 80\% of their final mass, as can be seen from the KS test in Fig.~\ref{fig:groupma}. On average $z_{80}=0.58\pm0.05$ for FG($R_{200}$) and $z_{80}=0.25\pm0.04$ for nFG($R_{200}$), with the KS test yielding $p_{\mathrm{KS}} = 6*10^{-6}$ on the $z_{80}$ distributions. A similar trend is also found for FG(0.5$R_{200}$) and nFG(0.5$R_{200}$) with $p_{\mathrm{KS}} = 3*10^{-2}$ for $z_{80}$ (Table~\ref{table:groupma}).
 
\begin{figure}
\centering
\includegraphics[width=1\linewidth]{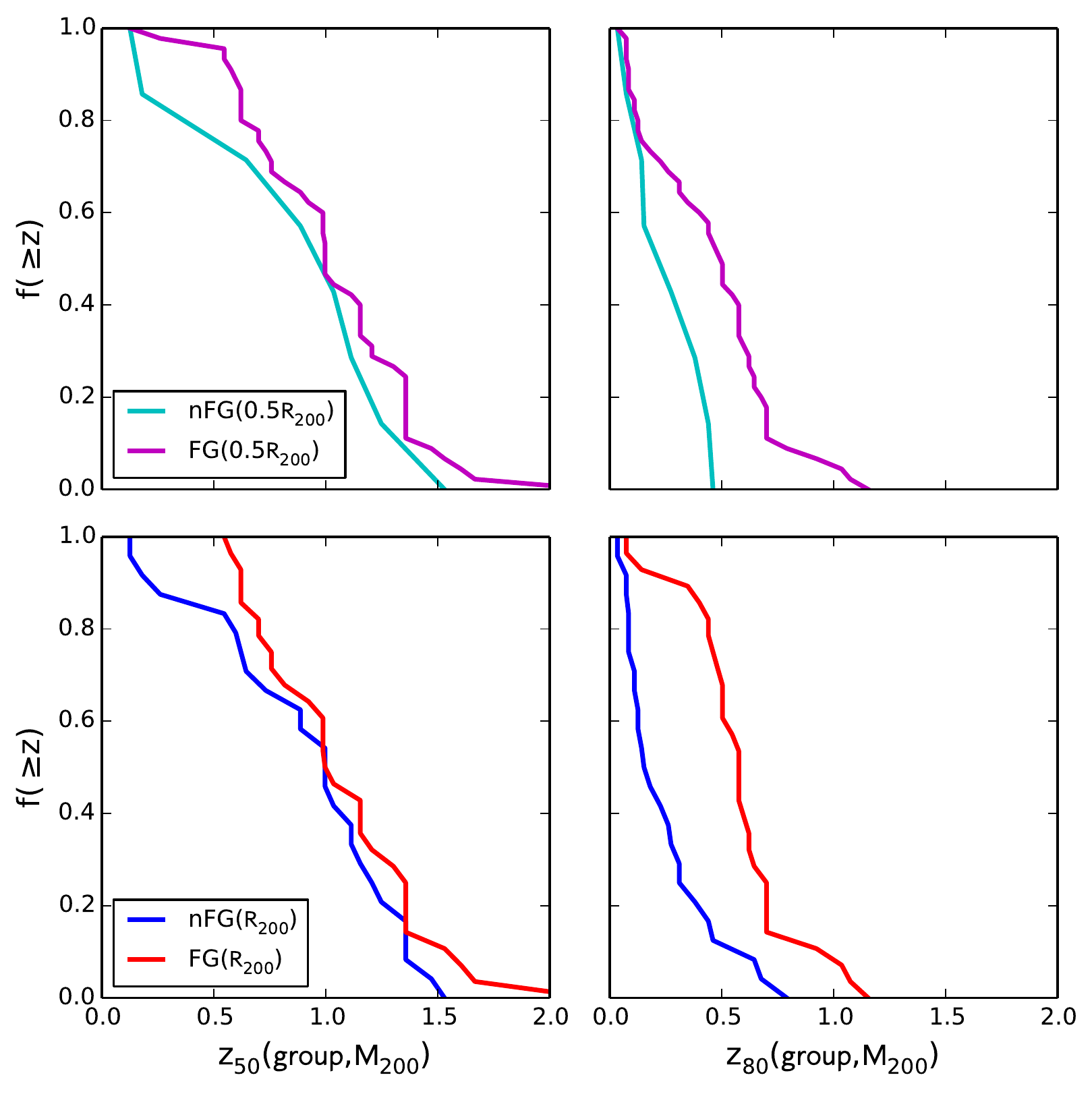} 
\caption{Distribution of group $z_{50}$ (\textit{left}) and $z_{80}$(\textit{right}) assembly times. Fossils and non-fossils are found to have a similar $z_{50}$, the redshift at which 50\% of the final $z=0$ $M_{200}$ mass is assembled. $z_{80}$, the redshift at which 80\% of the final $z=0$ halo mass is assembled, is significantly earlier for fossils than non-fossils, with most FG($R_{200}$) reaching $z_{80}$ before $z\sim0.4$.}
\label{fig:groupzf}
\end{figure}

In Fig.~\ref{fig:groupzf}, we show the cumulative distribution function of $z_{50}$, for comparison to previous fossil studies, and $z_{80}$, which we find in Illustris to be the most divergent mass assembly time for fossils and non-fossils. It is clear from this figure, that the steep increase in $M_{200}(z)/M_{200}(z=0)$ shown in Fig.~\ref{fig:groupma} for fossils between $z=0.4-0.8$, is a result of a large fraction of the fossil sample reaching $z_{80}$ in this window. In contrast, the distribution of non-fossil $z_{80}$ times extends over $z=0-0.8$ with the majority of non-fossils reaching $z_{80}$ between $z=0-0.4$. In general, we find $z\sim0.4$ to be a dividing line in the distribution of $z_{80}$ for FG($R_{200}$) and nFG($R_{200}$): $\sim$80\% of fossils reach $z_{80}$ before $z\sim0.4$, while $\sim80$\% of non-fossil reach $z_{80}$ after this time. Qualitatively, these results are also found with less significance for FG(0.5$R_{200}$) and nFG(0.5$R_{200}$). 
 
A lack of recent halo accretion, as indicated by an early $z_{80}$, may indeed be fundamental for the formation of a large gap within $R_{200}$ at the present day. In our mass regime, bright satellites originally were accreted as centrals of other groups, and their arrival is therefore associated with mass build up for the primary group. Thus an early $z_{80}$ ensures that no bright satellites fall in to the group, maintaining any gap that forms through mergers. While for non-fossils, the recent growth of their halos is related to the arrival of their brightest current satellites $m_{2}(z=0)$. For non-fossils, $z_{80}$ must occur within the last few Gyr such that accreted bright satellites do not have time yet to merge with the central, while $z_{80}$ must occur early enough for fossils such that any bright satellites accreted during this time of halo mass growth have had time to merge by the present day.
 
We also here note that although we only find a difference in the recent accretion history of fossils and non-fossils, we also find fossils are associated with overmassive BGGs (Fig.~\ref{fig:sr_bcg}), suggesting overmassive BGGs may also be associated with an early $z_{80}$ and not with $z_{50}$ as has been previously thought. As a direct check of the relation between halo assembly and the overmassiveness of the BGG, we select the most extreme 20\% of the distribution of $z_{50}$ times for our groups to check for association with \textit{f*}(BGG). By a two-sample KS test on the \textit{f*}(BGG) values of the earliest and latest forming halos we find \textit{f*}(BGG) for extreme early and late $z_{50}$ are not distinct ($p_{\mathrm{KS}}=0.7$). Comparatively, selecting the most extreme $z_{80}$ shows deviating distributions of \textit{f*}(BGG) ($p_{\mathrm{KS}}=0.01$), where groups with an early $z_{80}$ are associated with an overmassive BGG. Thus we truly find overmassive BGGs are associated with early $z_{80}$ and thus lack of recent group accretion, instead of an early halo formation time. This can also explain properties of the evolution in \textit{f*}(BGG)($z$) shown in Fig.~\ref{fig:m12z}. While the average \textit{f*}(BGG)($z$) of fossils does indeed grow due to mergers in the past few Gyr, it is equally important that the average \textit{f*}(BGG)($z$) of non-fossils has decreased over this time due to recent halo-halo mergers contributing to the group $M_{200}$.

In summary, we find the main difference in the mass assembly history of fossils and non-fossil groups is with respect to the recent accretion history, instead of an early formation time as has been found in other studies. Particularly we find fossils on average reach $z_{80}$ at an earlier epoch than non-fossils, indicating a lack of recent accretion. This difference in $z_{80}$ suggests a difference in the local environment of FG($z=0$) and nFG ($z=0$) groups over the past few Gyr, namely that present day fossil groups may exist in a relatively less dense environment which has prevented recent infall of new massive satellites. And indeed, this would be in agreement with previous studies of the environment in which fossil groups reside \citep{adami2007,dariush2010,cui2011,diazgimenez2011}.

\begin{table}
\centering
\caption{Group mass assembly times}
\begin{threeparttable}
\begin{tabular}{l c c}
\hline
Subsample & $z_{50}(\mathrm{group},M_{200})$ & $z_{80}(\mathrm{group},M_{200})$ \\ [0.5ex]
\hline
nFG(0.5$R_{200}$) & 0.85 $\pm$ 0.16 & 0.24 $\pm$ 0.05 \\
FG(0.5$R_{200}$) & 1.02 $\pm$ 0.06 & 0.45 $\pm$ 0.04 \\
\hline
nFG($R_{200}$) & 0.90 $\pm$ 0.08 & 0.25 $\pm$ 0.04 \\
FG($R_{200}$) & 1.07 $\pm$ 0.07 & 0.58 $\pm$ 0.05 \\
\hline
\end{tabular}
\tablecomments{Average redshift at which the groups assemble 50\% and 80\% of their final $M_{200}(z=0)$ mass. 1$\sigma$ errors have been bootstrapped.}
\label{table:groupma}
\end{threeparttable}
\end{table}

 
\section{Discussion}\label{sec:discussion}
 
\subsection{Comparison to other simulations}

The initial \citet{jones2003} formation scenario for the magnitude gap proposed a difference in the halo mass assembly history of fossil and non-fossil groups: the large magnitude gap of fossils formed as a result of early accreted massive satellites merging with the central galaxy, boosting the luminosity and mass of the central while depleting the bright end of the satellite population. Thus, testing how the magnitude gap relates to halo age has been of great interest to many theoretical fossil studies.
 
The mass assembly of fossil clusters ($M_{vir}\sim10^{14} M_{\odot}$) was first investigated by \citet{donghia2005}, and later the mass assembly of fossils in the low mass group regime ($M_{200}=10^{13}-10^{13.5} M_{\odot}/h$) was examined using the Millennium simulation \citep{dariush2007,dariush2010,gozaliasl2014b} as well as using other N-body cosmological simulations \citep{vonbendabeckmann2008,deason2013}.
 
When examining the full mass assembly history of group halos, \citet{dariush2007,dariush2010} and \citet{gozaliasl2014b} find fossils on average have assembled more of their final halo mass at nearly every redshift. Particularly they find the initial mass build up of fossils and non-fossils is different, including a difference in $z_{50}$ which is also returned by \citet{vonbendabeckmann2008} and \citet{deason2013}. While according to these simulations the halo formation time on average is earlier for fossil systems, there is also a considerable amount of scatter relating halo formation time and the magnitude gap of the group. Indeed many early forming systems are missed by selecting large magnitude gap groups and a non-significant amount of fossils also have a recent formation time \citep[e.g. see][]{dariush2010,deason2013,raouf2014}.
 
Our result of finding no difference in the $z_{50}$ of fossils and non-fossils in Illustris may then be related to the sample size of available Illustris groups in our selected mass regime compared to the much larger sample sizes of groups in the Millennium simulation used by the previous fossil studies. Given the large scatter previously reported in these studies, it is possible that by chance the distribution of $z_{50}$ for fossils is similar to the distribution for non-fossils in Illustris. The relative abundance of early-forming fossils will thus need to be examined in future larger cosmological simulations with hydrodynamics, such as the IllustrisTNG \citep{pillepich2017}. However, even with our smaller sample size, we find $z_{80}$ is significantly different for fossils and non-fossils in Illustris. Thus we expect that with larger sample sizes, fossil groups should show also a lack of recent accretion as important to the formation of the magnitude gap. Nevertheless, an earlier $z_{80}$, as we find, or an earlier $z_{50}$, as has been found in other studies, suggests fossil groups assemble some portion of their halo mass at an earlier epoch than non-fossils, following the original \citet{jones2003} idea.
 
We also find support for the idea of a `fossil phase' \citep{vonbendabeckmann2008} whereby groups only temporarily exist with a large magnitude gap due to recent mergers of satellites with the central without recent infall of new satellites, in good agreement with previous studies of simulated fossils \citep{vonbendabeckmann2008,dariush2010,gozaliasl2014b,kanagusuku2016}. As can be seen in our Fig.~\ref{fig:m12z},  we find that the large magnitude gap characterizing fossil groups at $z=0$ has only formed within the past few Gyrs.

Thus the picture of fossil group formation we find in Illustris relies on both the early accretion of massive satellites in addition to the lack of recent accretion of new bright satellites, both indicated by an early $z_{80}$. The early assembly of some fraction of a fossil group's halo mass allows enough time for $L*$ satellites to merge by the present day due to dynamical friction, producing a massive and luminous central galaxy. In combination, the lack of recent accretion for the halo ensures no new bright satellites replace those that have already merged and the gap formed through merging is preserved.
 
 
\subsection{Observational implications}
 
We find BGG properties consistent with what has been found in previous observational and theoretical studies, despite finding no difference in the $z_{50}$ group formation time of fossils and non-fossils. This may be because differences in group assembly have little effect on the assembly and properties of the BGG's stellar component; even large differences in the large-scale environment in which a halo forms do not show differences in BGG growth rate \citep{jung2014}, and differences in the the group halo formation time have not been found to produce observable differences in the stellar age or properties of their BGGs \citep{deason2013}.

We find no significant difference in the observational properties of fossil and non-fossil BGGs including color, sSFR, and stellar age. This is in good agreement with observational studies that find fossil BGGs seem to have typical properties of other ellipticals of the same mass, including the age and metallicity of stellar populations \citep{labarbera2009,harrison2012,eigenthaler2013} and number of globular clusters \citep{alamomartinez2012}. Additionally observed fossil BGGs follow the Fundamental Plane, Kormendy relation, and Faber-Jackson relation \citep{mendezabreu2012}.
 
There is also some recent evidence that fossil BGGs are not evolving passively, or at least not more passively than the BGGs of non-fossil groups. Evidence for recent fossil BGG activity includes: radio-loud AGN \citep{hess2012}, surface brightness profiles that deviate from a Sersic profile in the NIR \citep{mendezabreu2012} and optical \citep{alamomartinez2012}, apparent shell features \citep{eigenthaler2012}, unrelaxed X-ray isophotes \citep{miller2012}, tidal tails \citep{zarattini2016}, and ongoing merging in HST imaging \citep{ulmer2005}. This is in line with the more recent last major merger and recent significant growth of the BGG due to merging we find for fossils, and as has been found in other theoretical studies as well \citep[e.g.][]{diazgimenez2008,kanagusuku2016}.

Observations of the halo concentration parameter and X-ray scaling relations of fossil groups have often been interpreted with respect to the $z_{50}$ formation time found for fossils in previous simulation studies. Groups with an early $z_{50}$ are expected to have more concentrated halos for a given mass \citep[e.g.,][]{neto2007}, and early-forming groups have been speculated to follow different scaling relations \citep[e.g.,][]{jones2003,khosroshahi2007}. However, a wide range of concentration parameters have been measured for fossil groups \citep[e.g.,][]{khosroshahi2004, khosroshahi2006a, khosroshahi2007,democles2010,pratt2016}, and fossil groups seem to follow the same scaling relations as normal groups \citep[e.g.,][]{harrison2012,girardi2014,kundert2015}. These observations might be understood then if there is no difference in the $z_{50}$ of fossils and non-fossils, as we find here with Illustris. Furthermore, while we propose an early group $z_{80}$ is important for the development of the large magnitude gap of fossils at the present day, an early $z_{80}$, reflecting a lack of recent accretion, would be unlikely to affect the halo concentration or scaling relations.

 
\section{Summary and conclusions}\label{sec:conclusions}

We investigate the formation of the optical magnitude gap for galaxy groups with mass M$_{200}=10^{13}-10^{13.5} M_{\odot}/h$ in the Illustris cosmological simulation. Our analysis relies on studying the properties of fossil groups ($\Delta m_{12} \geq 2$) and non-fossils ($\Delta m_{12}=0-2$) defined by their gap within 0.5$R_{200}$ and $R_{200}$. The evolution of the groups is examined between $z=0-10$ with particular focus on the BGG stellar mass assembly and merger history, and assembly of the $M_{200}$ group mass. No significant difference between FG and nFG defined by their gap within 0.5$R_{200}$ is found, and thus we base our interpretation of the physical processes driving the formation of the gap on our analysis of FG and nFG defined by their gap within $R_{200}$.
 
Within $R_{200}$, approximately $\sim 0.4$ Mpc for our groups, the average gap of $\Delta m_{12}(R_{200}) \sim 1$ is consistent with Press-Schechter predictions. In agreement with observations, we find fossils have in general a more massive and more luminous central galaxy in comparison to non-fossils of the same group $M_{200}$, and additionally we find a significant correlation between the gap and the stellar mass of the BGG, implying both features are related and may have an origin due to the same process.
 
Our primary findings on the evolution of fossil group properties include:
 
\begin{itemize}
 
\item The magnitude gap, $\Delta m_{12} \geq 2$, of fossils identified at $z=0$ on average forms $\sim$ 3 Gyr ago, and is coincident with fossil BGGs becoming overmassive for their group $M_{200}$ mass compared to non-fossil BGGs on average. We furthermore find groups with a large magnitude gap at any redshift appear to also have a relatively more massive BGG than small magnitude gap groups.
 
\item Fossil BGGs become more massive than non-fossil BGGs due to increased mass acquired through mergers between $z=0.1-1$. Fossil BGGs are more likely to experience a greater number of major mergers, and more recently experience a major merger as compared to non-fossil BGGs. On average fossil BGGs have assembled $\sim 60$\% of their mass at $z=0$ from mergers, with the greatest contribution originating from major mergers.
 
\item While fossil BGGs both assemble 50\% of their final stellar mass and experience their last major merger $\sim$ 1 Gyr more recently than non-fossil BGGs, no difference is found in the observational properties of these BGGs including stellar ages, metallicities, and star formation rates.
 
\item The group mass assembly of fossils and non-fossils differs in only the recent group accretion history, particularly as indicated by differences in the distribution of $z_{80}$($M_{200}$) assembly times. $\sim80$\% of fossil groups reach $z_{80}$ before $z=0.4$, while $\sim80$\% of non-fossil groups reach $z_{80}$ after this epoch. Unlike studies of fossils in other simulations, we find no difference in the $z_{50}$($M_{200}$) of our groups, and in general no difference in the mass assembly histories of the groups at early times.
 
\end{itemize}
 
The primary difference between fossils and non-fossils is thus the mass assembly history of the group. The large magnitude gap and massive BGG of fossils is due to the merging of early arriving massive satellites, and lack of recent infall of new massive satellites over the past few Gyr. In Illustris, we find the magnitude gap of a group does not provide information on the dynamical state of the system, nor the age of the BGG, but instead seems primarily associated with the recent accretion history of the group within the past few Gyr.


\acknowledgments
 
AK received support from the Chandra-NASA Archive Grant AR6-17015X. ED gratefully acknowledges the support of the Alfred P. Sloan Foundation. JALA was funded by the MINECO (grant AYA2013-43188-P).
 
The Illustris simulations were run on the Odyssey cluster supported by the FAS Science Division Research Computing Group at Harvard University. We thank the Illustris collaboration for making their data and catalogs publicly available.
 
 {\footnotesize
\bibliographystyle{apj.bst}

}

\end{document}